\documentclass[english,pra,twocolumn,showpacs,superscriptaddress]{revtex4}
\usepackage[T1]{fontenc}
\usepackage[latin9]{inputenc}
\usepackage{amssymb}
\usepackage{graphicx}
\usepackage{esint}
\usepackage{verbatim}
\usepackage{babel}

\makeatletter

\newcommand{\oT}{\omega_\mathrm{T}}
\newcommand{\hoT}{\hbar \oT}
\newcommand{\kBT}{k_\mathrm{B} T}

\usepackage{babel}
\begin{document}

\bibliographystyle{apsrev}
\title{Collective excitations of a trapped Fermi gas at finite temperature}
\author{A. Korolyuk}
\author{J.J. Kinnunen}
\affiliation{COMP Centre of Excellence and Department of Applied Physics, Aalto University, FI-00076 Aalto, Finland}
\author{P. T\"orm\"a}
\email{paivi.torma@aalto.fi}
\affiliation{COMP Centre of Excellence and Department of Applied Physics, Aalto University, FI-00076 Aalto, Finland}
\affiliation{Kavli Institute for Theoretical Physics, University of California, Santa Barbara, California 93106-4030, USA}
{\abstract We study collective excitations of a trapped Fermi gas at finite temperature using the Bogoliubov-deGennes mean-field theory in conjunction with the generalized random phase approximation. The collective excitations are analyzed through the density response of a monopole excitation. We show the appearance of several modes at the normal fluid - superfluid phase transition temperature: Higgs mode-like pairing amplitude modulations, analogues of the second sound in the trapped gas, and an edge mode that is also the strongest mode in the response.}

\maketitle

\section{Introduction}

Collective excitations of many-body quantum systems provide a 
unique access for analyzing the effect of phase transitions and symmetry breaking.
Ultracold trapped atomic gases provide an ideal system 
for studying these phenomena: collective excitations at phase transitions 
have been studied in Fermi gases by tuning the interaction 
strength~\cite{Bartenstein2004a,Kinast2004a,Kinast2005a,Altmeyer2007a} 
and temperature~\cite{Sidorenkov}, and in Bose gases at the Mott insulator-superfluid phase transition~\cite{Endres2012}. For the bosonic gas the phase transition was observed to produce
a collective Higgs mode~\cite{Endres2012,Pollet2012},
but the mode has been observed also in superconductors~\cite{Sooryakumar1980a,Littlewood1982a}, and it is expected to be present also in superfluid Fermi 
%gases~\cite{Varma2001,Arovas2011,Varma2013,Tsuchiya2013}. 
gases~\cite{Varma2001,Tsuchiya2013}. 
Another mode unique for superfluids is the 
second sound~\cite{Lane1947,Taylor2009a,He2008a}, which involves the 
relative motion of the superfluid and the normal fluid components. 
The mode is well known in liquid helium~\cite{Lane1947} and it has been studied
experimentally in elongated trapped Fermi gases~\cite{Sidorenkov}.
And finally, two-band superconductors have been predicted to support a collective mode,
known as the Leggett mode~\cite{Leggett1966a}, describing the relative 
phase modulation of the superconducting order parameters of 
the two bands. Observation of the mode has been reported for a multiband MgB$_2$ superconductor~\cite{Blumberg2007a},
but not yet in other systems.

In this letter we study collective excitations in trapped Fermi gases. 
Theoretically, collective modes in ultracold Fermi gases are well 
described by the hydrodynamic 
%model~\cite{Griffin1997a,Baranov2000a,Bruun1999a,Stringari2004a,Bulgac2005a,Stringari2013a} 
model~\cite{Baranov2000a,Bulgac2005a,Stringari2008a} 
in the strongly interacting regime. However, in the more weakly interacting regime where the single particle character of the atoms becomes important, the generalized random phase approximation (GRPA) together with the
Bogoliubov-deGennes (BdG) mean-field theory provide a way for studying collective
phenomena. 
%%NEWNEWNEWNEW
While mean-field theories cannot grasp all the details of strongly interacting systems, the basic 
mean-field theory can be improved by including, for example, Gorkov-Melik-Barkhudarov induced interaction 
corrections~\cite{Gorkov1961a,Heiselberg2000a}. However, the qualitative picture remains the same and such corrections amount to renormalizing the underlying two-particle interactions. Indeed, mean-field theories have proven to be able to provide a good qualitative description of the relevant physics in atom gases at zero temperature~\cite{Stringari2008a}. 
Finite temperatures, and particularly the superfluid-normal phase transition are more challenging for mean-field based theories. Within simple mean-field theory in the normal state, the atoms interact only with the mean-field density and any non-trivial correlations in the normal state are lost. However, when calculating the response using GRPA, some of the 
neglected correlations are regained. Still, the results must be considered as only qualitative, particularly in more strongly interacting gases.
%%NEWNEWNEWNEW
The combined BdG + GRPA method has been used for studying various collective modes at 
zero~\cite{Bruun2002a,Ohashi2004,Korolyuk2011} 
and finite 
temperatures~\cite{Bruun2001a,Grasso2005a}.
Here we use the method for analysing monopole (angular momentum $L=0$) collective modes across the superfluid-normal phase transition and predict modes that are of Higgs and second sound type, as well as a prominent Leggett-mode like edge mode.

\section{Method}

The response of a many-body quantum system to a probing field coupling to the density degrees of freedom can be described by the density response function, which is defined in the linear response theory as
\begin{eqnarray}
\mathcal{A}(\mathbf{r},\mathbf{r}',\omega)=&\sum_{n}&\frac{\left\langle \phi_{0}\right|\hat{\rho}(\mathbf{r})\left|n\right\rangle \left\langle n\right|\hat{\rho}(\mathbf{r}')\left|\phi_{0}\right\rangle }{\hbar \omega-\left(E_{n}-E_{0}\right)} \nonumber \\
&-&\frac{\left\langle \phi_{0}\right|\hat{\rho}(\mathbf{r}')\left|n\right\rangle \left\langle n\right|\hat{\rho}(\mathbf{r})\left|\phi_{0}\right\rangle }{\hbar \omega+\left(E_{n}-E_{0}\right)},
\label{eq:response}
\end{eqnarray}
where $\left|n\right\rangle $ are the eigenstates of the many-body Hamiltonian
$\hat{H}$ and $E_{n}$ are the corresponding energy levels, i.e.
$\hat{H}\left|n\right\rangle =E_{n}\left|n\right\rangle $. Thus the poles in the 
density response function $\mathcal{A}(\mathbf{r},\mathbf{r}',\omega)$ 
reveal the excitation spectrum $E_n$ of the many-body system. 
%%For simplicity, we use $\hbar = k_\mathrm{B} = 1$.

\begin{figure}
\includegraphics[width=0.99\columnwidth]{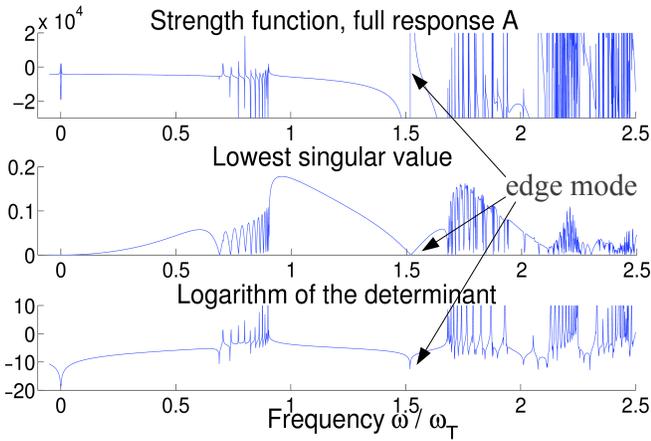}
\caption{Comparison of different methods for identifying resonant frequencies: (top) 
the strength function $S_{\uparrow\uparrow,L=0}\left(\omega\right)$,
(middle) the lowest singular value (LSV) of the matrix $1-K\left(\omega\right)$, and
(bottom) the determinant of the matrix $1-K(\omega)$. The response
is calculated as the response of the spin-$\uparrow$ density to the
probing field $h=\phi_{\uparrow}$.
Peaks of the strength function, zeroes of the LSV, and the
minima of the determinant indicate collective excitations. The edge 
mode (discussed later) is marked in the figure. Here $k_\mathrm{F}a = -0.56$.}
\label{fig:comparison}
\end{figure}

We study a two-component balanced atomic Fermi gas in a 3D spherically symmetric
trap using the BdG mean-field method. The system is described by
the Hamiltonian $\hat{H}=\hat{K}+\hat{U}$ with the single-particle Hamiltonian $\hat{K}=\sum_\alpha\int d\mathbf{r}\,\psi_{\alpha}^{\dagger}(\mathbf{r})\Bigl[\frac{-\hbar^2 \nabla^{2}}{2m}-\mu_\alpha+\frac{m\oT^2r^{2}}{2}\Bigr]\psi_{\alpha}(\mathbf{r})$, where $\alpha \in \{\uparrow,\downarrow\}$, describing atoms in the harmonic potential, and the interaction term $\hat{U}=-\int d\mathbf{r}\,\psi_{\uparrow}^{\dagger}(\mathbf{r})\psi_{\downarrow}^{\dagger}(\mathbf{r})\Delta\left(\mathbf{r}\right)+H.c.$, where the pairing field $\Delta\left(\mathbf{r}\right)= g_{0}\left\langle \psi_{\uparrow}(\mathbf{r})\psi_{\downarrow}(\mathbf{r})\right\rangle$ consists of different channels corresponding to the different angular quantum numbers $l$ of the atoms $\Delta(r) = \sum_l \Delta_l(\mathbf{r})$. The coupling constant $g_{0}$ is obtained from the s-wave scattering length $a$ as $\frac{1}{g_0}=\frac{m}{4\pi\hbar^2a}-\sum_{k}\frac{m}{\hbar^2 k^2}$, describing a contact interaction potential approximation for the two-body scattering T-matrix.

The Hartree energy (the term proportional to densities $n_{\uparrow\left(\downarrow\right)}(\mathbf{r})$) 
is known to be ill-behaved in the strongly interacting regime~\cite{Heiselberg2001a,Kinnunen2012a}, 
causing large perturbations in the density profiles by compressing the gas.
In BdG theory the problem
manifests already well before the actual divergence of the scattering length: the kinetic energy cost of adding one more atom to the center of the trap would be
\begin{equation}
   \delta E_\mathrm{kin} = \frac{\hbar^2 k_\mathrm{F}^2}{2m}.
\end{equation}
Assuming local density at the center of the trap of $n(0)$, the local Fermi momentum obeys $k_\mathrm{F}^3 = 3\pi^2 n(0)$
and the (Hartree) interaction energy gained equals
\begin{equation}
   \delta E_\mathrm{int} = gn(0) = \frac{4\pi\hbar^2 a}{m} n(0) = \frac{4\pi\hbar^2 a}{m} \frac{k_\mathrm{F}^3}{3\pi^2}.
\end{equation}
The problem with the Hartree shift appears when the interaction energy gain is larger than the kinetic
energy cost $\left|\delta E_{int} \right| > \delta E_{kin} $, which occurs for $\left|k_\mathrm{F}a\right| \geq \frac{3\pi}{8} \approx 1.18$ (the scattering length $a$
will have to be negative). Indeed, if this limit is reached, more and more atoms are piled at the center of the trap. This increases the local density $n(0)$
and hence the local Fermi momentum, violating the above limit even further and resulting in an unphysical collapse of the trapped gas.
Indeed, if one liked to include the Hartree shift, one would need to impose a cut-off energy or take 
into account the momentum dependence of the two-body scattering T-matrix. Both approaches would greatly increase
the complexity of the numerical calculation since the compression of the gas involves populating artificially
high energy states. Hence we neglect the Hartree energy but observe that this has also the positive side effect that it 
shows more clearly the effect of interactions eventually introduced by the GRPA for quasiparticle 
excitations in the normal state.
Notice that the problems with the Hartree shift ultimately derive from the attempt to describe interactions through the two-body scattering T-matrix
instead of the bare atom-atom interaction~\cite{Simonucci2011a}. These problems do not manifest for example in density functional theories which, when used
in the atomic gas context, do not separate the Hartree and exchange correlation energies~\cite{Bulgac2007a,Troyer2012a}.
%Including the Hartree shift in the weakly interacting regime does not 
%change the results qualitatively~\cite{Korolyuk2011}.

A probing field that couples to the density of spin-up atoms is described by an operator $\hat{V}=\int d\mathbf{r} \, \phi_{\uparrow}(\mathbf{r},t)\psi_{\uparrow}^\dagger(\mathbf{r}) \psi_{\uparrow}(\mathbf{r})$. Similarly one can calculate the density response for probing fields coupling to the density of spin-down atoms $\phi_\downarrow(\mathbf{r})\psi_{\downarrow}^\dagger(\mathbf{r}) \psi_{\downarrow}(\mathbf{r})$, and to the pairing field, $\eta(\mathbf{r})\psi_{\uparrow}(\mathbf{r}) \psi_{\downarrow}(\mathbf{r})$. Moreover, one could consider response to different linear combinations of the fields, for example the total density $\phi_\uparrow+\phi_\downarrow$ and the spin-response $\phi_\uparrow-\phi_\downarrow$ as in Ref.~\cite{Ohashi2004}. However, even though the choice of the probing field ($\phi_\uparrow$, $\phi_\downarrow$, $\eta$ or linear combinations of them) and the response field (up density, down density, total density, spin density, or the pairing field) do affect the actual coupling between the probing field and the various collective modes, the probed collective mode spectrum is the same and determined by the many-body interactions within the atom gas~\cite{Korolyuk2011}. Indeed, we have checked that the calculated collective mode spectrum is the same for all combinations although the visibility of different modes in the measured response do vary. We consider here mainly the response of the density of up-atoms to the probing field $\phi_\uparrow$, although we will also show the effect on the pairing field as some of the collective modes are dominantly pairing field excitations. Notice that we will calculate also directly the collective mode spectrum and that does not depend on the probing field. Furthermore, we assume the usual situation for ultracold Fermi gases where (pseudo)spins are fixed i.e. spin flips cannot occur.

In the GRPA method, the density response function can be calculated approximately using the mean-field Green's functions~\cite{CoteGriffin} $G_{ij}(\mathbf{1},\mathbf{2})=-\langle T\Psi_{i}(\mathbf{1})\Psi_{j}^{\dagger}(\mathbf{2})\rangle$, where $\Psi(\mathbf{1})=\left[ \psi_{\uparrow}(\mathbf{1}),\, \psi_{\downarrow}^{\dagger}(\mathbf{1})\right]^\mathrm{T}$ and $\mathbf{1}$ denotes a space-time 4-vector $\mathbf{x}_{1},t_{1}$. The response in our case can be written as a linear equation
%%\begin{widetext}
\begin{eqnarray}
&&A_{ij}(\mathbf{1},\mathbf{5})=A_{0ij}(\mathbf{1},\mathbf{5})+ \label{eq:equation_for_density_response_main}\\
&&g_{0}\sum_{k,l}\int d\mathbf{3} \, \left[L_{ik}^{kj}(\mathbf{1},\mathbf{3}) A_{ll}(\mathbf{3},\mathbf{5})- L_{ik}^{lj}(\mathbf{1},\mathbf{3}) A_{kl}(\mathbf{3},\mathbf{5}) \right], \nonumber
%%A_{ij}(\mathbf{1},\mathbf{5})=A_{0ij}(\mathbf{1},\mathbf{5})+g_{0}\sum_{k,l}\int d\mathbf{3} \, \left[L_{ikkj}(\mathbf{1},\mathbf{3}) A_{ll}(\mathbf{3},\mathbf{5})- L_{iklj}(\mathbf{1},\mathbf{3}) A_{kl}(\mathbf{3},\mathbf{5}) \right],
\end{eqnarray}
%%\end{widetext}
where $L_{ik}^{lj}(\mathbf{1},\mathbf{3})=G_{ik}(\mathbf{1},\mathbf{3})G_{lj}(\mathbf{3},\mathbf{1})$ and the indices $i,k,l,j \in \{\uparrow,\downarrow\}$. The function $A_{0}(\mathbf{1},\mathbf{5})=L_{i\uparrow}^{\uparrow j}(\mathbf{1},\mathbf{5})$ is called the single-particle response. 
%We consider here only spherically symmetric monopole modes, i.e. the probing field will not provide any angular momentum to the atoms.

%%We calculate the mean-field Green's functions using the Bogoliubov-deGennes method (BdG). The method amounts to a mean-field approximation in which the the two-particle interaction is replaced by the mean-field interaction operator $\hat{U}_\mathrm{mf}=-\int d\mathbf{r}\left(\psi_{\uparrow}^{\dagger}(\mathbf{r})\psi_{\downarrow}^{\dagger}(\mathbf{r})\Delta\left(\mathbf{r}\right)+H.c.\right)$, where $\Delta\left(\mathbf{r}\right)=\left|g_{0}\right|\left\langle \psi_{\uparrow}(\mathbf{r})\psi_{\downarrow}(\mathbf{r})\right\rangle $ is the order parameter (gap). We neglect Hartree energy (term proportional
%%to densities $n_{\uparrow\left(\downarrow\right)}(\mathbf{r})$) in order to 
%%more clearly show the effect of beyond mean-field interactions introduced
%%by the GRPA for excitations in the normal state. This provides also a great
%%simplification for the computation since the Hartree shift is ill-behaved
%%in the strongly interacting regime~\cite{SOMEONE}. Including the Hartree 
%%shift in the weakly interacting regime would not change the results 
%%qualitatively~\cite{Korolyuk2011}.

How strongly the probing field couples to a collective excitation can be calculated
using the strength function ~\cite{Grasso2005a}, which for the monopole mode is defined as
$S_{ij}\left(\omega\right)=\int dr_{1}dr_{5}\,r_{1}^{4}r_{5}^{4} A_{ij}(r_{1},r_{5},\omega)$,
where $A_{ij}(r_1,r_5,\omega)$ is obtained from the response $A_{ij}(\mathbf{1},\mathbf{5})$ by the Fourier
transform. This is the dynamic structure factor response to a spherically symmetric modulation of the trapping potential with frequency $\omega$. Positions of the poles in the strength function yield the energies of the collective excitations and the strength of each pole provides information on the coupling between the particular excitation and the probing field. However, the linear form of Eq.~(\ref{eq:equation_for_density_response_main}) provides
also alternative ways for analyzing the collective excitations. The equation has the form $A=A_{0}+KA$ 
which satisfies a formal solution $A=\left(1-K\right)^{-1}A_{0}$,
where $A$ and $A_{0}$ are the full and the single-particle responses,
respectively. Collective excitations in $A$ appear as zeroes of the 
determinant $\det\left(1-K\right)$. Alternatively, one can use the singular value decomposition $1-K=U\Sigma V^{\dagger}$,
where $\Sigma$ is a diagonal matrix whose diagonal elements are the
sought-for singular values of the matrix $1-K$, and $U$, $V$ are unitary
matrices, allowing one to write the equation in the form $A=V\Sigma^{-1}U^{\dagger}A_{0}$.
If the lowest singular value (LSV) is zero, the response $A$
diverges revealing a collective excitation. 
%%Thus we have three numerical 
%%approaches for evaluating the collective modes of the system: the strength
%%function $S_{L=0}(\omega)$, the singular value decomposition of 
%%the matrix $1-K$ and the determinant $\det\left(1-K\right)$.

Fig.~\ref{fig:comparison} shows that all three methods give
the same excitation spectrum. One should notice, however, that the 
determinant and the singular
value decomposition neglect the contribution from the single-particle
response $A_0$ in $A=\left(1-K\right)^{-1}A_{0}$. This is important in cases where the probing field
does not couple with the system ($A_0 = 0$), and 
this will be relevant in the superfluid-normal phase transition as will
be seen below. However, since LSV gives 
generally the best resolution, we will use it for determining the 
frequencies of the collective modes. On the other hand, the 
2x2 matrix structure of the strength function with elements $S_{ij}$ provides an additional
view to the nature of the collective modes. The geometric sum of the diagonal elements of the matrix 
$S_\rho = \sqrt{S_{\uparrow,\uparrow}^2 + S_{\downarrow,\downarrow}^2}$ describes the response in 
the densities and the sum of the off-diagonal elements $S_\Delta = \sqrt{S_{\uparrow, \downarrow}^2 + S_{\downarrow, \uparrow}^2}$
describes excitations in the pairing field. The relative strengths of the density and pairing
field contributions of each mode can be thus easily accessed by calculating the gap/density ratio
$R=\frac{S_\Delta^{2}}{S_\Delta^{2}+S_\rho^{2}}$. Figs.~\ref{fig:t00},~\ref{fig:t02} and~\ref{fig:g10} show through colour coding the gap/density ratio $R$ along the 
collective modes. 

\section{Density response spectra}

\begin{figure}
\includegraphics[width=0.99\columnwidth]{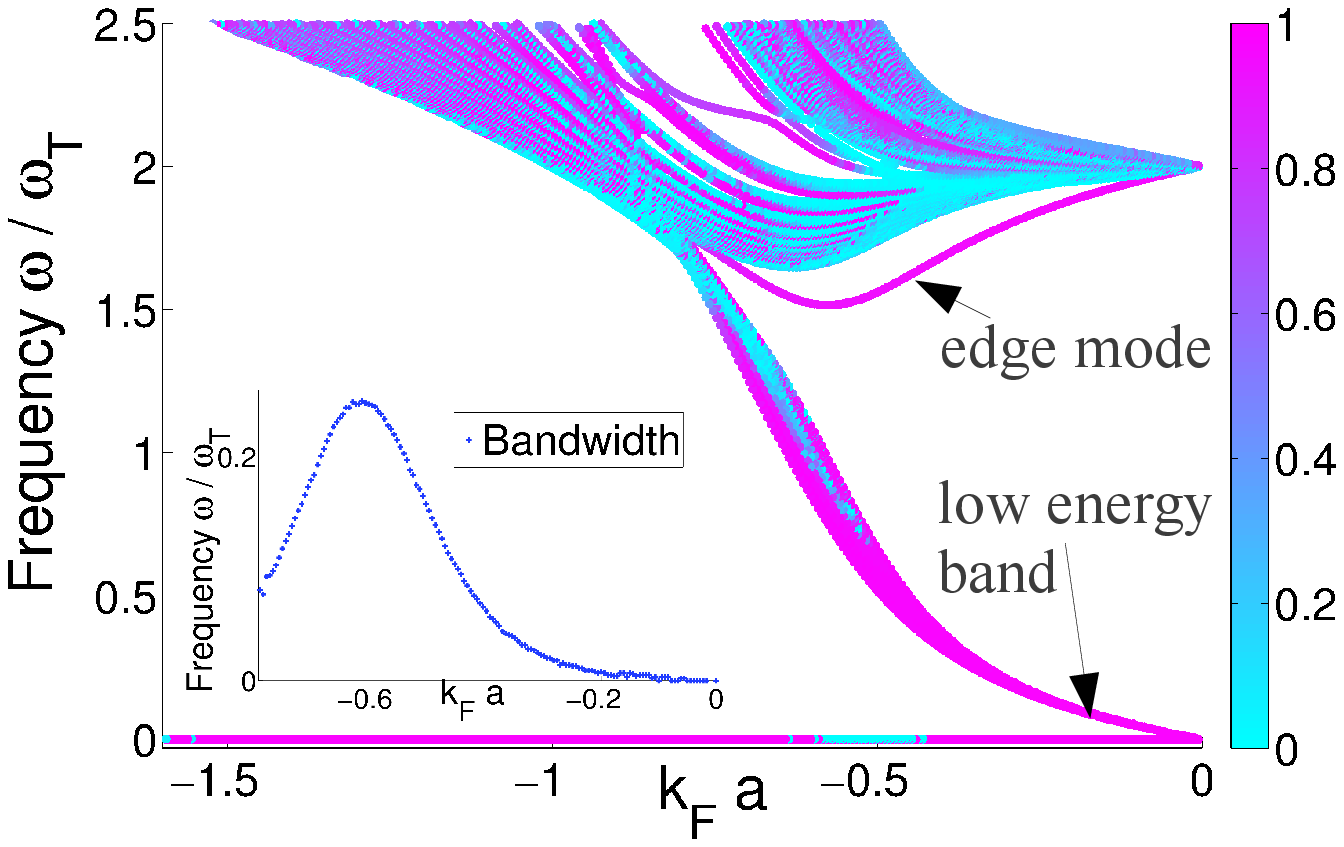}
\includegraphics[width=0.99\columnwidth]{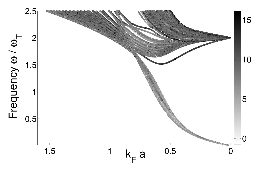}
\caption{(Color online) Top: collective mode frequencies at zero temperature (obtained from LSV) as a function of the interaction strength $k_{F}a$. Marked in the figure are the edge mode and the low-energy branch. Inset shows the bandwidth of the low-energy branch.
The color coding describes gap/density ratio $R$: blue for $R = 0$, red for $R=1$, see main text. Bottom: the same collective modes obtained from the strength function $S (\omega)$, which is the response to a modulation of the trapping potential with the frequency $\omega$. The color codes show the logarithm of the height of the peak in the density response.}
\label{fig:t00}
\end{figure}

Fig.~\ref{fig:t00} shows the collective excitations at zero temperature. 
The collective mode along the $(\omega=0)$-axis is a signature of a Goldstone mode that has been shown to describe phase fluctuations of the order parameter $\Delta\left(r\right)$~\cite{Ohashi2004}. 
The figure reveals also a very prominent lone excitation, labeled 'edge mode', that in the weakly interacting regime lies between the two bands of excitations starting from $\omega=0$ and $\omega=2\,\oT$. 
The edge mode bears resemblance to the Leggett mode predicted for two-band superconductors and it is also the collective mode located farthest from the center of the trap, as discussed below in Section~\ref{sec:leggett}.
At weak interactions the low energy band can be identified as Higgs mode-like pair vibration modes with the energies of the excitations scaling as $\hbar \omega \sim 2\Delta(r=0)$~\cite{Littlewood1982a,Bruun2002a}, where $\Delta(r=0)$ is the pairing gap at the center of the trap. The inset in Fig.~\ref{fig:t00} shows the bandwidth of the low-energy band. 
Notice the narrowing near the point where the low energy band merges with the higher energy excitations. 

\begin{figure}
\includegraphics[width=0.99\columnwidth]{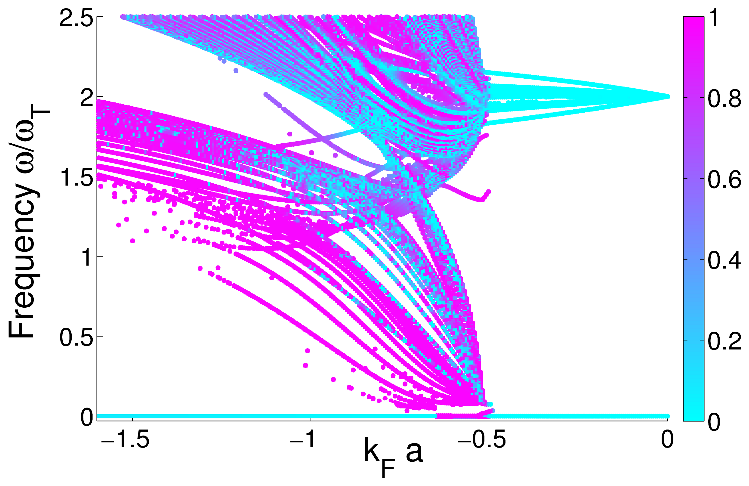} %%t02_v1.eps}
\includegraphics[width=0.99\columnwidth]{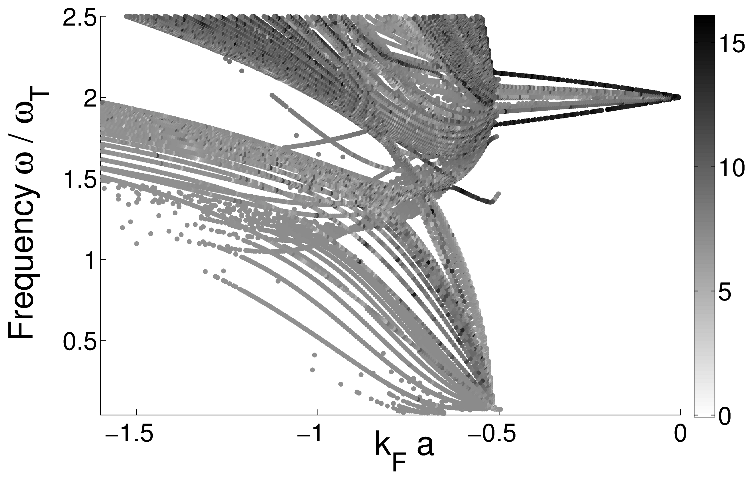}
\caption{(Color online) Top: collective excitations calculated with LSV as a function of the interaction strength at temperature $\kBT =0.2\,\hoT$. The superfluid-normal state phase transition occurs at $k_\mathrm{F}a = -0.52$. Bottom: collective excitations calculated from the strength function. The color coding is as in Fig.~\ref{fig:t00}.}
\label{fig:t02}
\end{figure}

Fig.~\ref{fig:t02} shows the collective modes for a finite temperature $\kBT =0.2\,\hoT$ calculated using 
both the singular value and the strength function.
For interactions weaker than critical (for this temperature $\left(k_{F}a\right)_\mathrm{crit}\approx-0.52$)
the pairing gap in the gas is equal to zero, $\Delta(r)=0$, and the gas is in the normal state.
This implies that, since our model neglects the Hartree energy shift, the mean-field BdG
Hamiltonian is effectively non-interacting. Still, the collective excitations 
shown in the figure for the normal state are not equal to $\omega = 2\,\oT$, as
one would expect for a non-interacting gas. Indeed, the energy shift is a result
of the interactions between the quasiparticles introduced by
the GRPA. The magnitude of the edge mode is proportional to the order parameter and thus it vanishes in the vicinity of the phase transition.

\begin{figure}
\includegraphics[width=0.99\columnwidth]{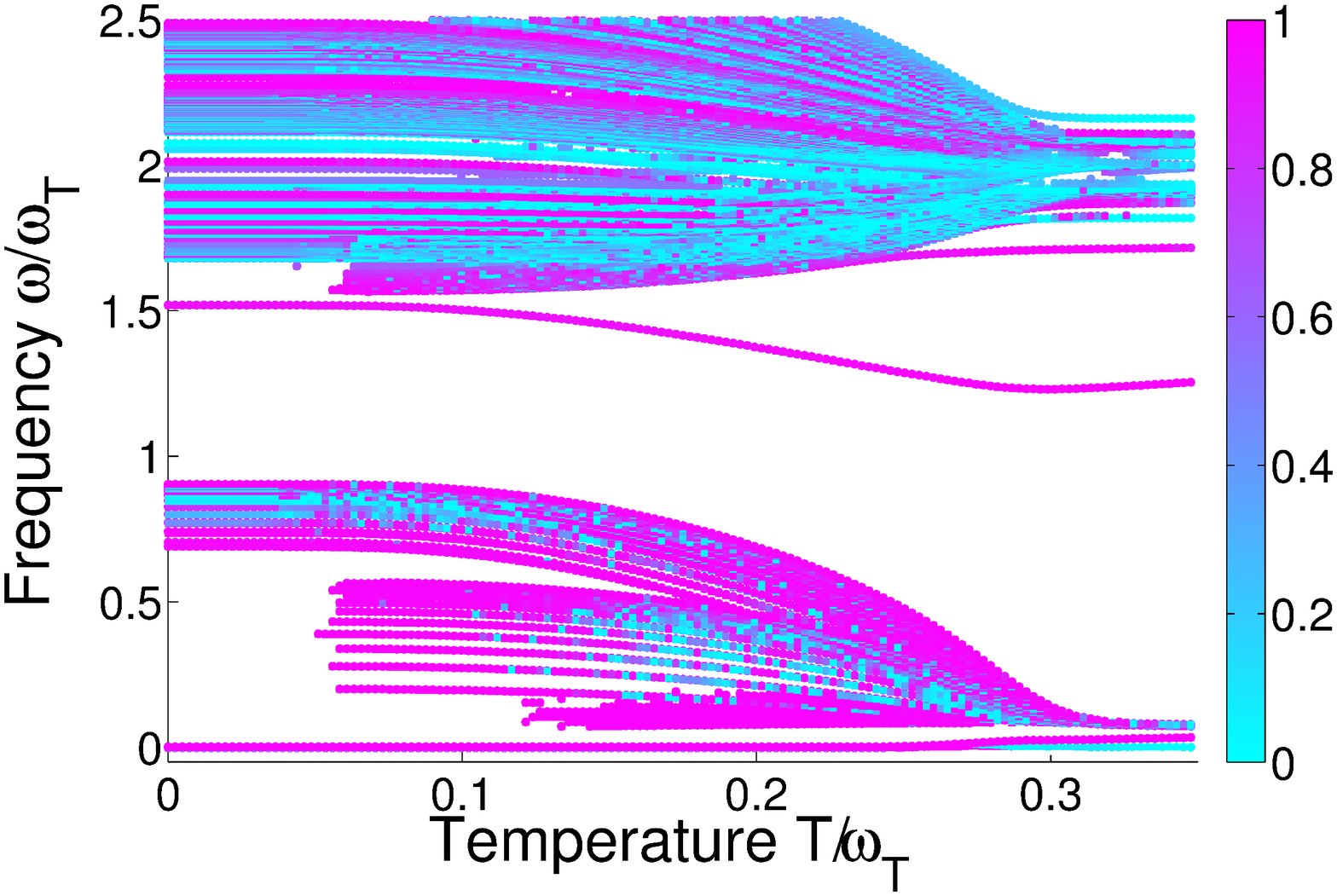}
\includegraphics[width=0.99\columnwidth]{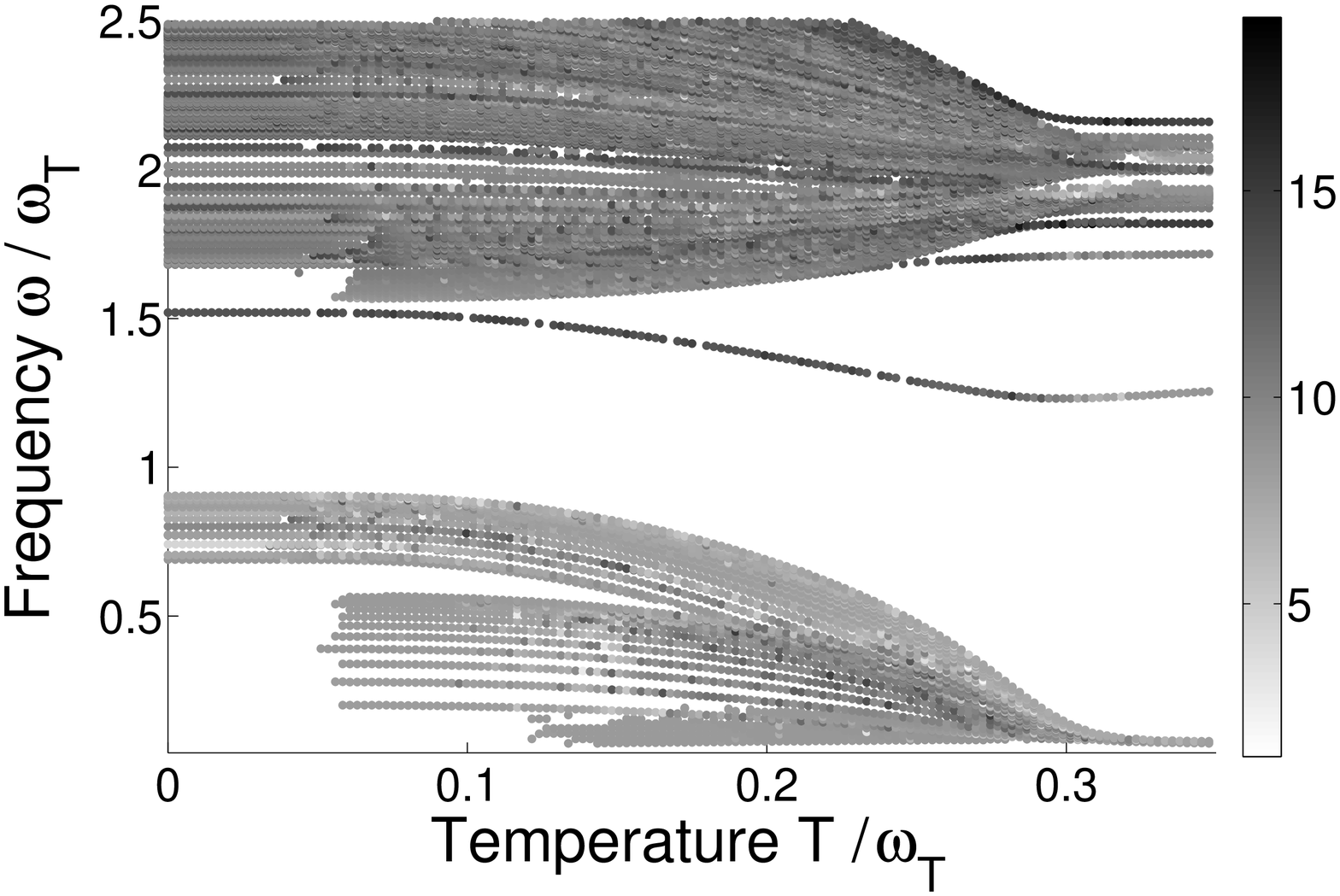}
\caption{(Color online) Top: collective excitation frequencies calculated with LSV as a function of temperature for the interaction strength $k_{F}a=-0.56$ (corresponding to $\Delta\left(r=0\right)=1.22\,\hoT$ at zero temperature). Bottom: collective excitations calculated from the strength function. The color coding is as in Fig.~\ref{fig:t00}}
\label{fig:g10}
\end{figure}

Fig.~\ref{fig:t02} shows also that at finite temperatures new modes emerge. 
These are more readily seen in Fig.\ref{fig:g10}, where the excitations are plotted as a function
of temperature for a fixed interaction strength $k_\mathrm{F}a=-0.56$. The figure shows 
the emergence of two branches of excitations at temperature $\kBT \approx 0.07\,\hoT$,
one branch at around $\omega \approx 1.6\,\oT$ and 
%one at $\omega<0.5\,\oT$.
%The latter branch has a distinctive second sound-like nature. 
one branch with a distinctive second sound-like nature at $\omega < 0.5\,\oT$.
The second sound describes relative motion of the normal and superfluid components requiring the presence of both components. 
Here the branch appears with thermal excitations and vanishes when the superfluid component vanishes around the phase transition at $\kBT \approx 0.3 \,\hoT$.
The question whether the modes correspond to relative motion of normal and superfluid components is, however, beyond the scope of the present model. 
Notice that the edge and pair vibration modes remain artificially visible even in the normal phase in Fig. 4. The difference with Fig.~\ref{fig:t02} is that the 
phase transition obtained from the BdG is not as sharp when increasing the temperature compared to tuning the interaction strength. These are numerical
artifacts partly due to finite size effects (the discreteness of the single-particle spectrum) and partly due to the slow convergence of the BdG iteration when the order
parameter becomes very small. This results in visible but very weak residual edge and pair-vibration modes in the plot even in the normal state (the magnitudes
of the modes across the phase transition drop by at least two orders of magnitude).

\section{Collective mode analysis}

\begin{figure}
\includegraphics[width=0.99\columnwidth]{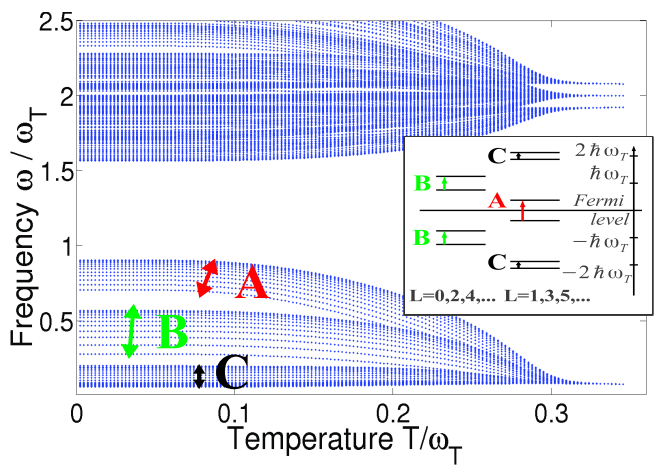}
\caption{Quasiparticle level transition energies $E_{jl}-E_{j'l}$ for $k_\mathrm{F}a=-0.56$. 
Three low-energy branches (A, B, C) are marked on the plot. Inset: quasiparticle energy level structure and the corresponding transitions A,B, and C.}
\label{fig:energy_levels}
\end{figure}

Most of the collective modes in Fig. \ref{fig:g10} can be understood by analyzing the quasiparticle energy levels $E_{jl}$ obtained from the BdG method.
Fig.~\ref{fig:energy_levels} shows the low energy transition energies $E_{jl}-E_{j'l}$ as a function of temperature. 
However, not all modes in Fig.~\ref{fig:g10} have corresponding quasiparticle level transitions, the edge mode being the most prominent.
Furthermore, although one can quite easily associate most of the collective modes in Fig.~\ref{fig:g10} with quasiparticle transitions in Fig.~\ref{fig:energy_levels}, the matching is not quite exact as the actual collective modes have slightly lower frequencies. 
This is again the effect of remnant quasiparticle interactions recovered by GRPA. 

Fig.~\ref{fig:energy_levels} (right) describes
how the branches A, B and C arise from the transitions between energy levels, providing a 
simple interpretation for many of the collective modes in Fig.~\ref{fig:g10}. Only the branch A appears at zero temperature as the branches B and C are blocked by the Pauli exclusion principle. 
However, the blocking effect is relaxed at finite temperatures when thermal quasiparticle 
excitations appear. These are the microscopic counterparts of the normal fluid component, and 
the $B$ and $C$ transitions can be understood as the propagation of the normal fluid excitation e.g. a hole within the superfluid. On the other hand, the same transition can be understood also as 
the reverse propagation of a bound Cooper pair, revealing the dual nature of the second sound-like band in Fig.~\ref{fig:g10}. 
The transition lines within each branch A, B, and C are also ordered: the smaller the angular momentum
$l$ of the quasiparticle state, the higher the transition energy. This can be
understood by observing that higher angular momenta $l$ correspond
to wave functions located further away from the trap center. Hence the
effective gap experienced by the quasiparticle level is lower (as the gap decreases
monotonically towards the edge) and consequently the transition energy is lower.

\section{Collective mode profiles}

\label{sec:leggett}

The density response $A_{ij}(\mathbf{1},\mathbf{5})$ defined in Eq.(2) allows one to study the spatial dependence of the various collective modes. The strength function $S(\omega)$ provides a measure for the strength of the coupling between the probing field and each collective mode, and it has thus direct experimental relevance. However, in order to 
obtain better understanding on the nature of various modes, and especially the prominent lone mode that we call the edge mode, one can study the spatial dependence of the density response $A_{ij}(r_1,r_5,\omega)$ in more detail.

\begin{figure}[h!]
\includegraphics[width=0.49\columnwidth]{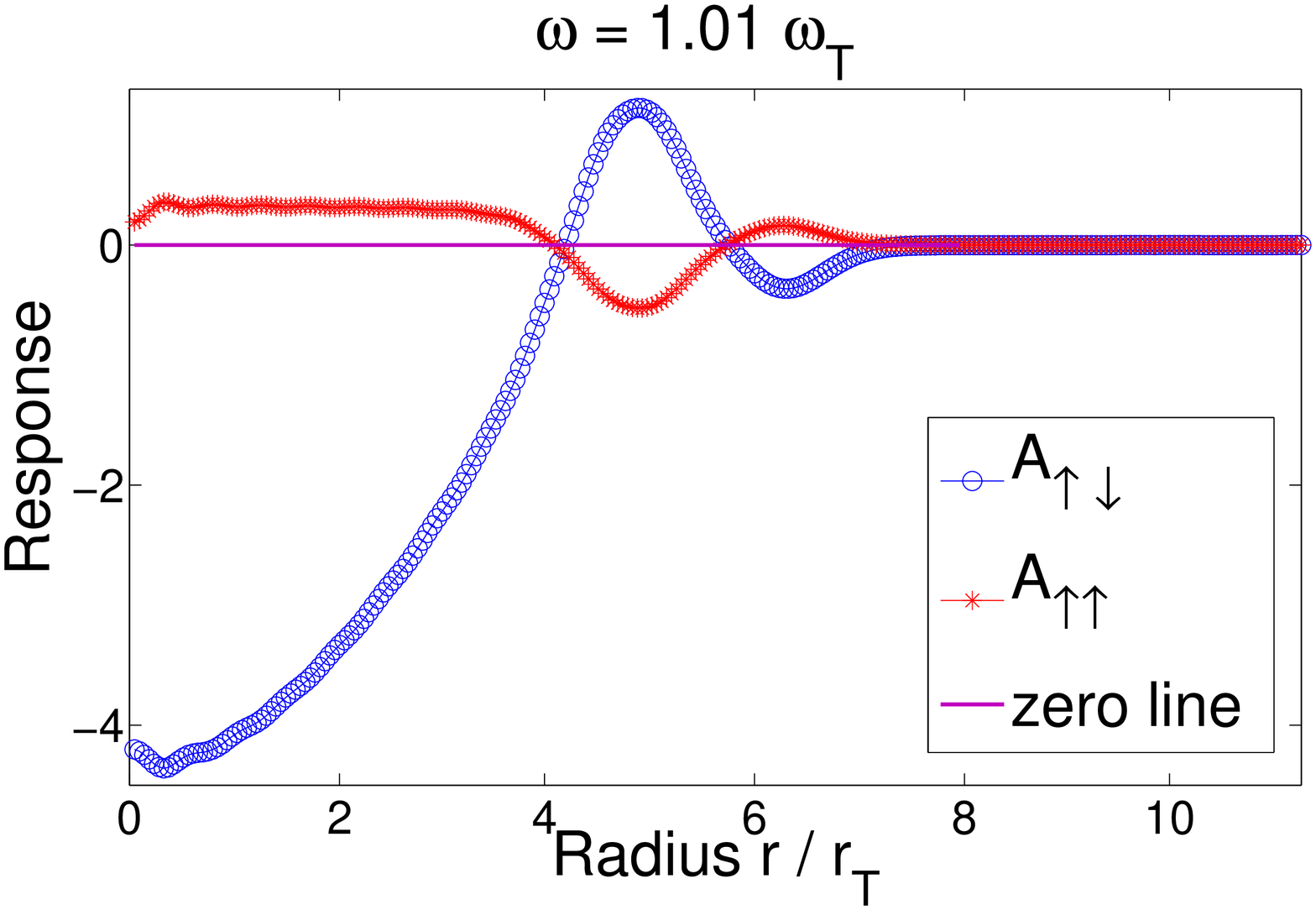}
\includegraphics[width=0.49\columnwidth]{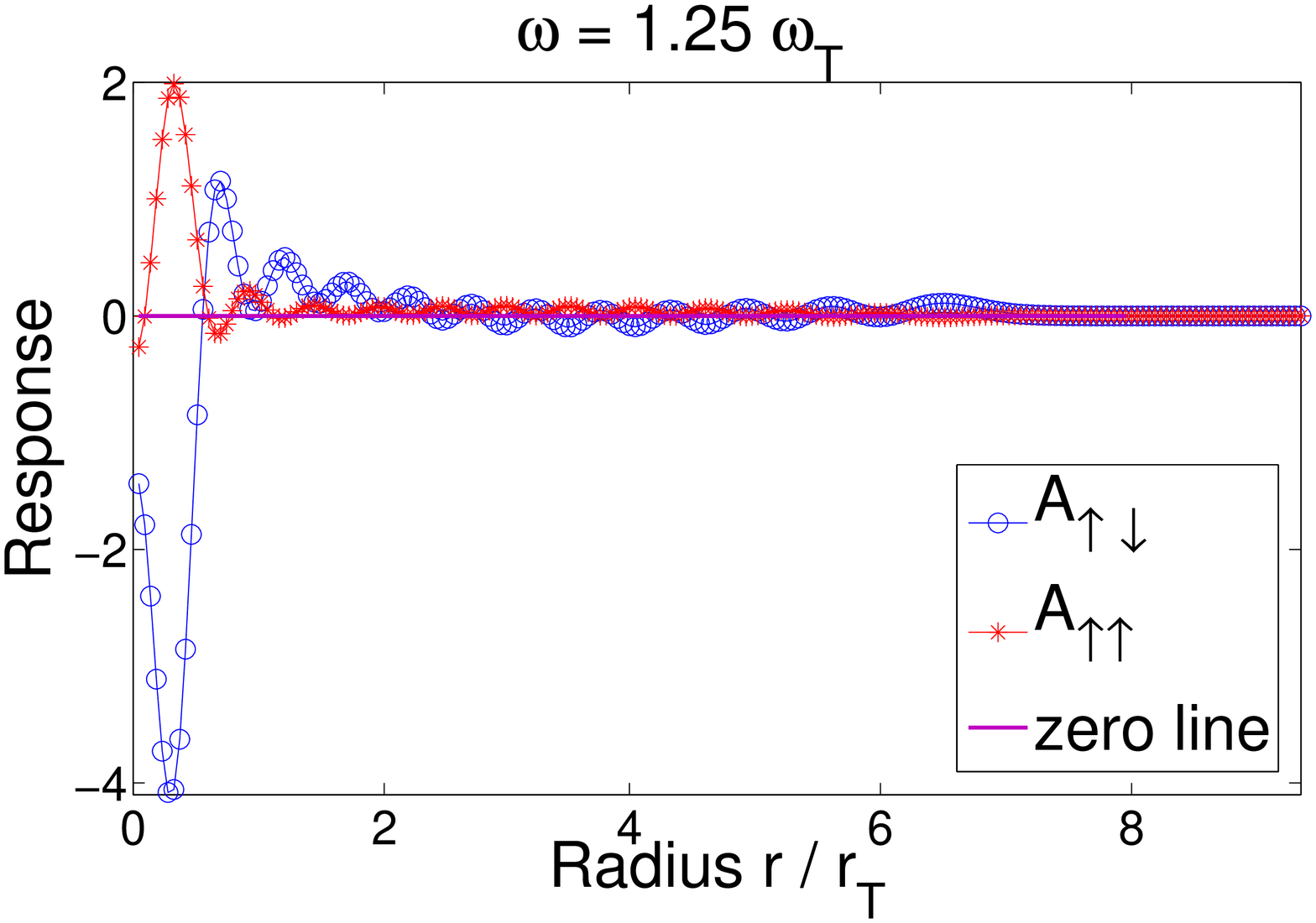}
\includegraphics[width=0.49\columnwidth]{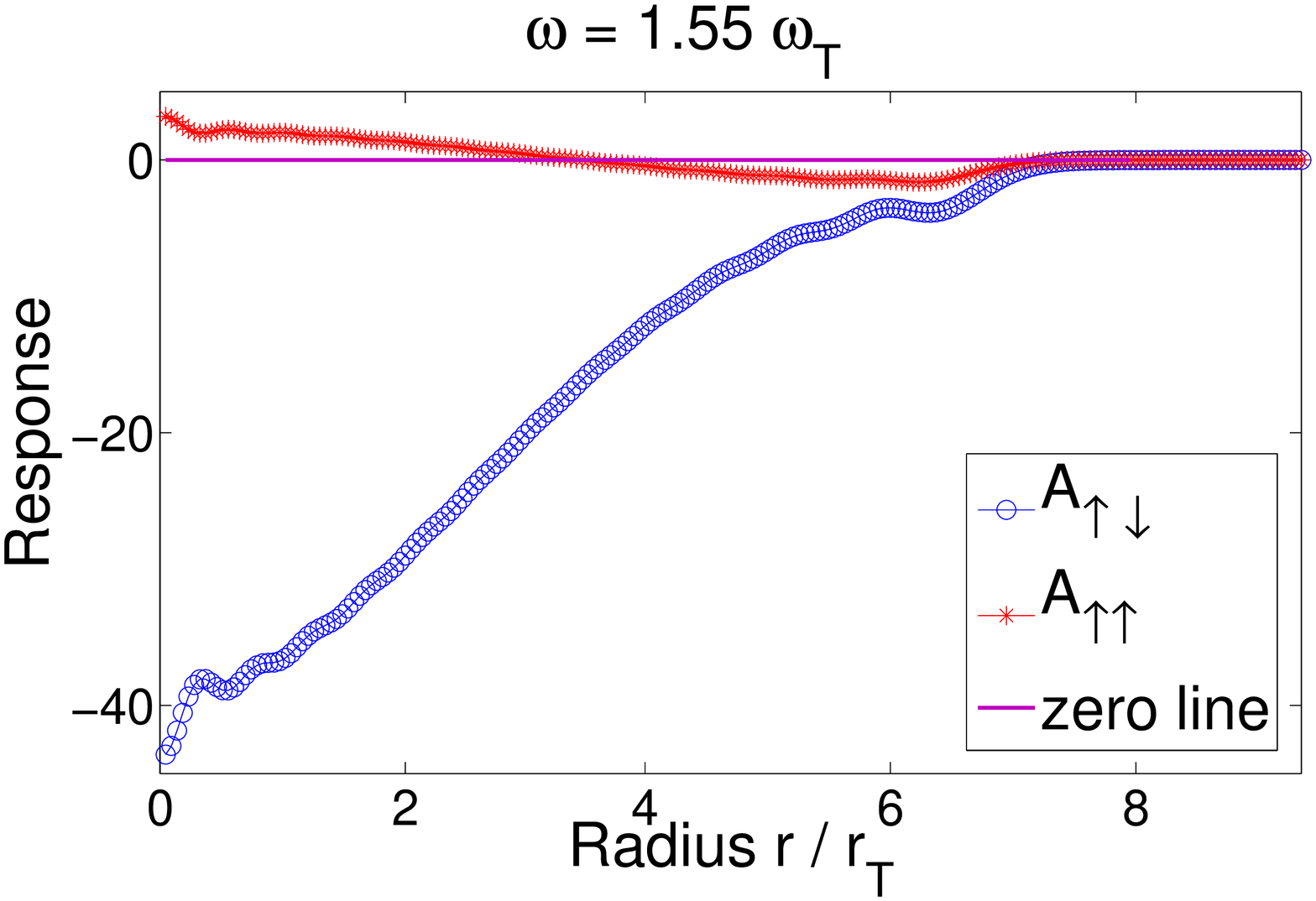}
\includegraphics[width=0.49\columnwidth]{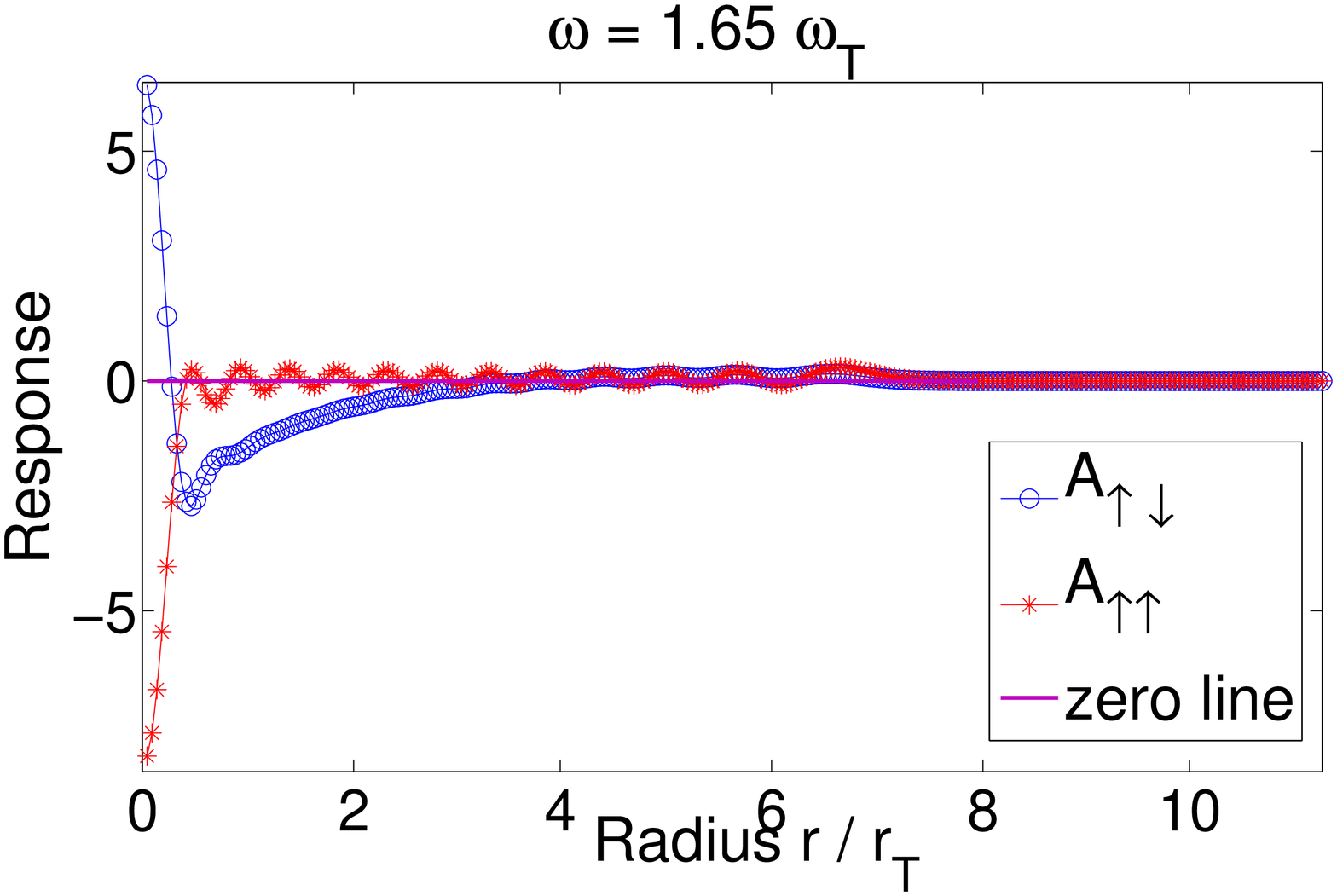}
\caption{The integrated density response $A_{ij}(r,\omega)$ as a function of position $r$ calculated for two modes in the low energy branch ($\omega = 1.01\,\oT$ and $\omega=1.25\,\oT$), for the edge mode ($\omega = 1.55\,\oT$) and for one mode from the upper branch ($\omega=1.65\,\oT$), see also Figure~\ref{fig:edgerad}. Here  $k_\mathrm{F}a = -0.56$ and temperature $T = 0$. Shown are the pairing field $A_{\uparrow \downarrow}$ and the spin-up density $A_{\uparrow \uparrow}$ components of the 2x2 matrix response. Here $r_\mathrm{T}=\sqrt \frac{\hbar}{m \oT}$ is oscillator length.}
\label{fig:edgewf}
\end{figure}

\begin{figure}
\includegraphics[width=\columnwidth]{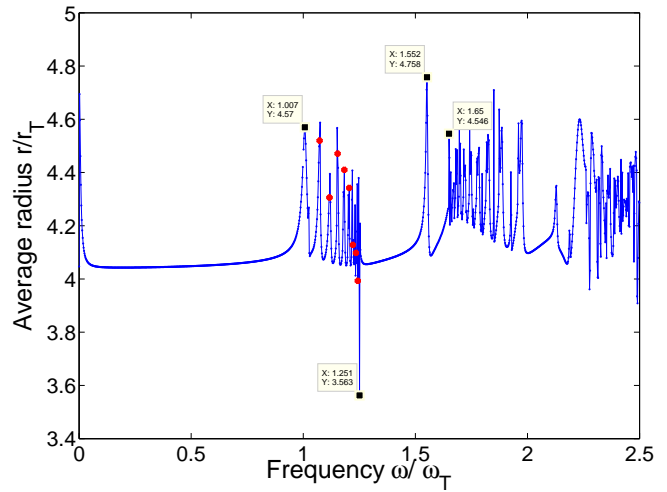}
\caption{The average radius of the collective mode profile $A_{\uparrow \downarrow}(r_1,r_5,\omega)$ as a function of frequency $\omega$. Only the values of $\omega$ corresponding to actual collective modes are physically relevant (shown with red circles for the low energy band). Four of the modes marked in the figure correspond to the modes shown in Fig.~\ref{fig:edgewf}. The edge mode is the mode with the highest average radius, i.e. it is the mode located farthest from the center of the trap.}
\label{fig:edgerad}
\end{figure}

Figure~\ref{fig:edgewf} shows four collective mode profiles 
calculated by integrating the density response over the radius $r_5$ i.e.
\begin{equation}
  A_{ij}(r,\omega) = \int dr_5\, r_5^2 A_{ij}(r_1,r_5,\omega).
\end{equation}
The frequency $\omega$ is chosen to match the corresponding collective mode frequency: the figures with $\omega = 1.01\,\oT$ and $\omega = 1.25\,\oT$ correspond to the two extremal collective modes (highest and lowest frequencies, respectively) within the band of Higgs-like excitations shown in Fig.~\ref{fig:t00}. 
The figure with $\omega = 1.55\,\oT$ corresponds to the edge mode and $\omega= 1.65\,\oT$ is the next higher frequency mode from the edge mode. 
For clarity these modes are marked also in Fig.~\ref{fig:edgerad}. 
Figure~\ref{fig:edgewf} shows that the collective modes are spread over the whole atom cloud. 
In addition, except for the edge mode, all other modes have nodes where the pairing field response component $A_{\uparrow \downarrow}(r,\omega)$ vanishes at some radius $r$. 
These nodes reflect the forms of the corresponding single-particle wavefunctions: 
for example, analyzing energies like in Fig.~\ref{fig:energy_levels}, we see that the single-particle transition corresponding to the $\omega=1.01\,\oT$ mode is for an atom with angular momentum quantum number $l=23$. Since the low energy transition is necessarily in the vicinity of the Fermi surface (the Fermi energy here is $E_\mathrm{F} = 24.5\,\oT$), the transition will need to elevate the atom from the occupied shell $n=0$ to the first unoccupied shell $n=1$. Indeed, the harmonic trap eigenstate for $l=23$, $n=1$ has a node near the point where the collective mode profile has a node, i.e. at $r = \sqrt{24.5}r_\mathrm{T} \approx 4.95 r_\mathrm{T}$. Similarly, higher energy modes in the 'Higgs' branch describe transitions of atoms with lower angular quantum numbers $l$ (all have odd $l$ since the states with even $l$ yield transitions in the higher energy band). Since the energy of the single-particle state is $E_{nl} = \oT(2n+l+3/2)$, in order for these lower angular momentum states to be at the Fermi surface, the shell quantum number $n$ is necessarily increased. Thus, the single-particle states involved in these transitions have a higher number of nodes, and this property is visible in the collective mode profiles, as exemplified by the $\omega=1.25\,\oT$ profile in Fig.~\ref{fig:edgewf}. 

However, comparing with the corresponding profiles of other modes, the edge mode has a striking absence of nodes in the pairing field response profile in Fig.~\ref{fig:edgewf}. We interpret this as a collective mode that does not influence the shell quantum number $n$ at all, but instead the transitions are between different $l$-quantum numbers only. The mode can be understood as a transition between different pairing field channels $\Delta_l(r)$: the total order parameter can be expressed as $\Delta(r) = \Delta_{l=0}(r) + \Delta_{l=1}(r) + \ldots$, where
each channel $\Delta_l(r)$ is physically distinct due to the angular momentum conservation for single particles. The mode is thus a Leggett mode-like collective excitation as it describes transitions between different pairing field channels, here the even and odd $l$-quantum number channels $\Delta_l(r)$. On the other hand, since the single particle transitions are required to conserve the angular momentum $l$ there are no corresponding single-particle excitations. The edge mode is thus a purely collective excitation and, like the Leggett mode, it can be understood as a Josephson-like oscillation between the different $l$-bands of the order parameter $\Delta_l$. 

Figure~\ref{fig:edgerad} shows the average radii of different modes, defined as
\begin{equation}
   \langle r\rangle(\omega) = \frac{\int dr_1 \int dr_5 \,r_1^3 r_5^2 |A_{\uparrow \downarrow}(r_1,r_5,\omega)|}{\int dr_1 \int dr_5 \,r_1^2 r_5^2 |A_{\uparrow \downarrow}(r_1,r_5,\omega)|}.
\end{equation}
While the average radius can be calculated for any frequency, the quantity has little meaning except when the frequency is on resonance with some collective mode excitation. The resonant frequencies, and the corresponding average radii,
are shown in the figure ~\ref{fig:edgerad} . The figure ~\ref{fig:edgerad}  shows that the edge mode is the one closest
to the cloud edge, hence the name 'edge mode'. This result is easily understood
by noticing that the nodes in the collective mode profiles, as shown in Fig.~\ref{fig:edgewf}, are located close to the cloud edge; the more modulations there are in the collective mode profile, the more weight the mode has in the inside of the cloud. 
Still, even for the edge mode, the actual cloud edge is still rather far (at approximately $6\,r_\mathrm{T}$). This is due to broad profiles of all low energy collective modes as seen in Fig.~\ref{fig:edgewf}.

\section{Experimental considerations}

The gap/density ratio $R$ displayed in Figs.~\ref{fig:t00},~\ref{fig:t02} and~\ref{fig:g10} show that the edge mode and the low energy band in the weakly interacting regime involve mainly gap modulations, as do the second-sound like low energy 
thermal excitations in the finite temperature results. 
%%This implies that these modes do not involve prominent density modulations and hence the associated modes
%%may prove difficult to observe experimentally. 
The gap, or pairing field, modulations might become more visible
by ramping the system across the BCS-BEC crossover, thus mapping the pairing field onto a condensate of molecules. 
The gap modulations in the BCS side would then become modulations in the molecular densities~\cite{Greiner2003a}. One could also consider the possibility that trap modulation at the collective mode frequency, being an efficient way of coupling energy to the system, could lead to increased heating due to possible coupling of the modes to other excitation. Increased heating could then be a way to observe those modes that are primarily gap, not density modes.
%%The density 
%%response can also be accessed by using Bragg spectroscopy that provides a momentum kick for the atoms~\cite{Vale2008a}. Combining this with a photon counting technique~\cite{Cornell2010a} would make visible modes 
%%that couple only weakly to the center of mass movement of atoms.
%%However, a quantitative analysis of such probing schemes would require going beyond the monopole mode and spherical symmetry.

All the results shown in this manuscript were obtained for a spherically symmetric gas of $4930$ atoms in each spin component.
As was discussed in Ref.~\cite{Korolyuk2011} the key energy scale is provided by the trapping frequency $\hoT$. This implies
that when scaling to larger systems, the interaction strength $k_\mathrm{F}a$ needs to be reduced in order to keep the ratio $\Delta(r=0)/\hoT$ constant. The interaction strength can be easily tuned experimentally, and the regime $\Delta(r=0) \sim \hoT$ has already been explored in several collective mode 
%experiments~\cite{Bartenstein2004a,Kinast2004a,Kinast2005a,Altmeyer2007a}. 
experiments~\cite{Bartenstein2004a,Kinast2004a,Kinast2005a}. 
In order to have the 
temperature in the same trapping frequency energy scale $\kBT \sim \hoT$, 
the experiments utilize elongated systems in which the radial ($\omega_\perp$) and axial ($\omega_{||}$) oscillator frequencies can differ by an order of magnitude, allowing the hierarchy of energy scales $\omega_{||} \ll \kBT \ll \omega_\perp$. While the results considered here assume spherical symmetry, one expects qualitatively similar features to be present in the two-dimensional-like radial modes in the perpendicular direction~\cite{Bartenstein2004a}. In contrast, the modes in the axial direction are more one-dimensional~\cite{Sidorenkov} and thus more different from the collective modes studied here.

\section{Conclusions}

To conclude, we have studied collective modes in finite temperature trapped Fermi gases with spherical symmetry. We have identified several collective modes appearing at the superfluid phase transition: second sound-like modes that disappear when the normal fluid component vanishes at zero temperature, and Higgs mode-like excitations describing modulations of the pairing field amplitude with little effect in the actual atomic densities. We also predict a striking edge mode that is the strongest mode in the response, which opens up a possibility to study Leggett-mode related physics in a controllable and microscopically simple system.

\section*{Acknowledgements}
This work was supported by the Academy of Finland through its Centers of 
Excellence Programme (2012-2017) and under Projects Nos. 135000, 141039, 251748, and 263347. This research was supported in part by the
National Science Foundation under Grant No. NSF PHY11-25915. Computing resources
were provided by CSC-the Finnish IT Centre for Science.

%\begin{thebibliography}{26}
%\end{thebibliography}

\end{document}